\documentclass{article}

\PassOptionsToPackage{numbers,sort&compress}{natbib}
\usepackage[preprint]{neurips_2026}

\usepackage[utf8]{inputenc}
\usepackage[T1]{fontenc}
\usepackage{amsmath}
\usepackage{amssymb}
\usepackage{array}
\usepackage{booktabs}
\usepackage{enumitem}
\usepackage{float}
\usepackage{graphicx}
\usepackage{microtype}
\usepackage[dvipsnames]{xcolor}
\usepackage{hyperref}

\newcommand{\papername}{Ablating Safety}
\newcommand{\method}{\textsc{Ablating Safety}}
\newcommand{\dataset}{\textsc{Security-AR}}
\newcolumntype{L}[1]{>{\raggedright\arraybackslash}p{#1}}

\title{Ablating Safety: Mechanisms for Removing Alignment in Language Models for Security Applications}

\author{
Isaac\\
University College London\\
\And
Arthur Gervais\\
University College London\\
}

\begin{document}
\raggedbottom

\maketitle

\begin{abstract}
Safety-aligned language models often refuse cybersecurity requests whose wording resembles misuse, even when the task is authorized and defensive. This makes security evaluation ambiguous: a failed answer may reflect missing capability or refusal-policy intervention. \method{} studies alignment removal as a controlled transformation-evaluation protocol for authorized security tasks, comparing authorized-context prompting, reversible refusal-direction activation projection, representation-control projections, and LoRA-based de-alignment or task adaptation. We evaluate refusal, attempt rate, validated security success, general-capability retention, instability, and out-of-scope unsafe compliance on \dataset{}, a 60-prompt suite of authorized security, benign general, and non-operational spillover probes. The reported runs include a four-model projection pilot with 416 completions, a three-model Qwen2.5 LoRA extension with 1,980 held-out completions, representation and robustness sweeps, and executable secure-repair validators. Single-vector refusal projection raises mean security score only from 0.46 to 0.50 while increasing unsafe compliance from 0.10 to 0.47; rank-4 refusal-subspace projection reaches 0.51 while matching the aligned spillover rate. Task-only LoRA raises mean security score to 0.87 with general score 0.83 and unsafe compliance 0.13, while refusal-suppression with retention raises spillover to 0.27. These results support evaluating alignment removal as a utility-risk frontier, not as an uncensoring recipe, and treating compliance alone as neither competence nor safe deployment.
\end{abstract}

\section{Introduction}

Language models are increasingly used in security workflows that require adversarial reasoning: vulnerability triage, binary reverse engineering, exploitability assessment, secure patch analysis, and controlled red-team exercises. These workflows are legitimate when they operate on authorized systems, isolated challenge targets, or defensive audit tasks. They also overlap lexically and procedurally with prompts that safety-aligned assistants are trained to refuse. As a result, refusal behavior can become a measurement confounder: an aligned model may fail a security benchmark because it refuses the task, not because it lacks the underlying capability.

Existing evaluations often treat this behavior as a property of the deployed assistant. A model either complies, refuses, or can be induced to comply with a jailbreak prompt. Recent mechanistic work suggests a different view: refusal can be mediated by low-dimensional activation directions in some open-weight models \citep{arditi2024refusal}, and practitioner workflows such as abliteration apply related transformations to suppress refusal behavior without full retraining \citep{labonne2024abliteration}. These observations raise a scientific question that is distinct from prompt jailbreak research: when alignment is removed through model-level transformations, what changes besides refusal?

This paper studies alignment removal as a controlled model transformation. The goal is not to release an ``uncensored'' model or to optimize harmful behavior. The goal is to measure whether safety behavior is mechanistically separable from task competence, and to identify which removal methods preserve useful reasoning while avoiding broad unsafe spillover. This framing matters for security evaluation because prompt-level bypasses, model edits, and fine-tuning changes are not equivalent interventions. They may expose different capabilities, induce different failure modes, and require different governance controls.

We ask:
\begin{quote}
\emph{Can refusal behavior be reduced for authorized security tasks while preserving general capability and bounding unsafe spillover outside the evaluation scope?}
\end{quote}

\method{} answers this question as an evaluation protocol and controlled empirical study. In this paper, ``ablating safety'' means applying a controlled transformation that weakens one measurable safety behavior---refusal---and then auditing what else changes. The evidence is organized in two layers. A projection pilot instantiates the prompt-baseline and reversible activation-projection setting across four open-weight models. An extended evaluation then exercises the protocol's fine-tuning-based arm with LoRA adapters on three Qwen2.5 models, using the same task suite, scoring rules, and responsible-release constraints.

The paper makes three contributions:
\begin{itemize}[leftmargin=*]
    \item We formalize alignment removal as a transformation-evaluation problem, with metrics and controls that separate refusal reduction from task success, general capability retention, and unsafe spillover.
    \item We define \dataset{}, a controlled benchmark and release design for authorized security tasks, including redacted public artifacts and decision rules for interpreting utility-risk tradeoffs.
    \item We implement a projection pilot and a LoRA extension showing that prompt, activation, representation-control, and fine-tuning interventions induce distinct utility-risk tradeoffs.
\end{itemize}

The paper is scoped to open-weight models and controlled security evaluations. It does not evaluate attacks on deployed third-party systems, does not recommend releasing transformed checkpoints, and does not publish operational exploit payloads. The empirical claims are bounded to the reported open-weight models, task suite, transformation families, and scoring rules; they are not claims that the observed frontier generalizes to frontier-scale systems or to all de-alignment methods.

\section{Background and Related Work}
\label{sec:background}

\paragraph{Guardrails, refusal, and safety alignment.}
Modern assistants build on foundation models whose broad capabilities also create broad misuse surfaces \citep{brown2020language,bommasani2021opportunities,touvron2023llama,jiang2023mistral}. Instruction-tuned assistants are commonly trained to follow benign requests while refusing unsafe or policy-sensitive requests \citep{ouyang2022training,bai2022constitutional}. Deployment systems also add external guardrails through programmable dialogue policies, input-output classifiers, and agent monitors \citep{rebedea2023nemo,inan2023llamaguard,chennabasappa2025llamafirewall}. These layers are important in practice, but they differ from model-internal refusal. A runtime filter can block or rewrite an output without changing the model's computation; safety fine-tuning changes the distribution the model itself samples from. This paper focuses on the latter: transformations that alter refusal behavior inside an open-weight model or its inference pass.

Refusal is useful for deployment but problematic as a capability-measurement primitive. A failed answer on a security benchmark may indicate lack of competence, successful safety enforcement, over-refusal, guardrail-triggered truncation, or a task-template artifact. Recent work argues that safety behavior can be shallowly expressed in early response tokens \citep{qi2024fewtokensdeep}, and that fine-tuning can erode safety even when the fine-tuning data are benign \citep{qi2023finetuning}. Follow-on work proposes constraining safety-token behavior during fine-tuning as a way to prevent alignment drift \citep{wang2026fewtokens}. These results motivate evaluating task competence and refusal policy as separate outcomes.

\paragraph{Representation engineering and activation steering.}
Representation engineering treats high-level model behaviors as directions or subspaces in activation space \citep{zou2023representation}. Activation addition and contrastive activation addition show that inference-time activation edits can steer behaviors without full optimization \citep{turner2023activation,panickssery2024caa}. This family of methods is directly relevant to alignment removal because it supplies both the measurement tools for finding behavioral directions and the intervention tools for adding, subtracting, projecting, or gating them.

\paragraph{Mechanistic refusal directions and circuits.}
Prior work finds that refusal in multiple open-weight chat models is mediated by a low-dimensional residual-stream direction: adding the direction can induce refusal, while erasing it can suppress refusal \citep{arditi2024refusal}. Later work generalizes and complicates this picture. Affine concept editing combines projection and activation addition to control refusal more precisely across models where pure projection can produce incoherence \citep{marshall2024refusal}. COSMIC automates refusal-direction selection without assuming output-level refusal templates \citep{siu2025cosmic}. Sparse-autoencoder steering identifies refusal-related features but also reports capability tradeoffs, suggesting that refusal-relevant features may be partly entangled with ordinary language-model competence \citep{obrien2024sae}. Recent preprints further argue that refusal should be understood as routing from detected concepts to output policy, not merely as a detectable concept vector or a surface refusal string \citep{frank2026routing,frank2026alignmentroutes}. We treat these papers as mechanistic motivation rather than settled consensus: they suggest candidate interventions and controls that must be validated under task, retention, and spillover metrics.

\paragraph{Abliteration and de-aligned models.}
Public ``abliteration'' workflows apply representation-level modifications to reduce refusal behavior in open-weight models \citep{labonne2024abliteration}. The term is used inconsistently in practitioner settings, covering activation projection during inference, weight orthogonalization, and downstream uncensored model releases. Abliteration-specific studies have begun to measure robustness and failure modes: extended-refusal training can make models more resistant to single-direction abliteration \citep{abushairah2025defense}, safety-pretraining checkpoints differ in how much refusal survives activation edits \citep{agnihotri2025granular}, and contrast-baseline construction can determine whether extracted directions are functional \citep{petrov2026topicmatched}. This paper contributes a security-evaluation view: alignment removal should be judged not by refusal rate alone, but by the utility-retention-spillover frontier it induces.

\paragraph{Jailbreaks and prompt-level bypasses.}
Jailbreak and red-teaming work studies prompts that cause aligned models to violate intended safety policies \citep{perez2022redteaming,ganguli2022redteaming,wei2023jailbroken,zou2023universal,chao2023jailbreaking,mazeika2024harmbench,chao2024jailbreakbench}. These attacks provide important context because practitioners often use prompt workarounds when aligned models over-refuse. However, prompt-level bypass and model-level removal are different interventions. A jailbreak changes the input context; ablation, fine-tuning, or activation intervention changes the model or its hidden-state trajectory. The reported implementation therefore uses an authorized-context prompt baseline to separate input framing from model-level transformation; stronger jailbreak baselines remain outside the public release.

\paragraph{Cybersecurity and capability evaluation.}
Cybersecurity benchmarks increasingly measure both defensive coding behavior and offensive-capability risk. CyberSecEval evaluates insecure-code generation and compliance with cyberattack requests \citep{bhatt2023cyberseceval}; Cybench specifies professional-level CTF tasks in executable environments and reports agent performance across models and scaffolds \citep{zhang2024cybench}. Code-generation and security studies show that model outputs can be useful while still introducing vulnerabilities \citep{chen2021evaluating,austin2021program,pearce2022asleep}. General retention can be measured with standard language, reasoning, and code evaluations such as MMLU, BIG-bench, TruthfulQA, and HumanEval-style tasks \citep{hendrycks2021measuring,srivastava2023beyond,lin2021truthfulqa,chen2021evaluating}. These benchmarks motivate \dataset{}, but the present paper adopts a narrower release boundary: authorized, sandboxed tasks for measuring security competence, with transformed checkpoints and live-target exploit procedures withheld from the public package.

\section{Problem Statement}
\label{sec:problem}

Let $M$ be an instruction-tuned language model that maps a prompt $x$ to a distribution over completions $y$. Let $\mathcal{X}_{sec}$ denote authorized security prompts, $\mathcal{X}_{gen}$ denote benign general-capability prompts, and $\mathcal{X}_{out}$ denote out-of-scope harmful prompts that are not legitimate evaluation tasks. A transformation $T$ maps $M$ to either a modified model $T(M)$ or an inference-time wrapper around $M$.

We evaluate $T$ using five quantities:
\begin{align}
    R(T) &= \mathbb{E}_{x \in \mathcal{X}_{sec}}[\mathrm{Refusal}(T(M), x)],\\
    A(T) &= \mathbb{E}_{x \in \mathcal{X}_{sec}}[\mathrm{Attempt}(T(M), x)],\\
    S(T) &= \mathbb{E}_{x \in \mathcal{X}_{sec}}[\mathrm{TaskScore}(T(M), x)],\\
    G(T) &= \mathbb{E}_{x \in \mathcal{X}_{gen}}[\mathrm{GeneralScore}(T(M), x)],\\
    H(T) &= \mathbb{E}_{x \in \mathcal{X}_{out}}[\mathrm{UnsafeCompliance}(T(M), x)].
\end{align}

The measurement problem is to estimate the Pareto frontier between lower refusal on authorized security tasks, higher validated task success, preserved general capability, and bounded unsafe compliance outside the task scope. Attempt rate is reported separately because non-refusal alone is not success: a model can stop refusing while producing useless, incoherent, or unsafe content. A transformation is useful only if it improves authorized task performance without collapsing general reasoning or broadly increasing unsafe behavior.

\section{The \method{} Protocol}
\label{sec:method}

\subsection{Overview}

\method{} evaluates all transformations through the same pipeline:
\begin{enumerate}[leftmargin=*]
    \item Select open-weight instruction-tuned models and base-model counterparts when available.
    \item Apply one alignment-removal transformation or a baseline condition.
    \item Run each condition on security, general-capability, and out-of-scope probe suites.
    \item Score outputs with sandbox validators, rubric-based review, and refusal classifiers.
    \item Report utility-risk confidence intervals over prompts, seeds, and decoding settings.
\end{enumerate}

\begin{figure}[t]
    \centering
    \includegraphics[width=\linewidth]{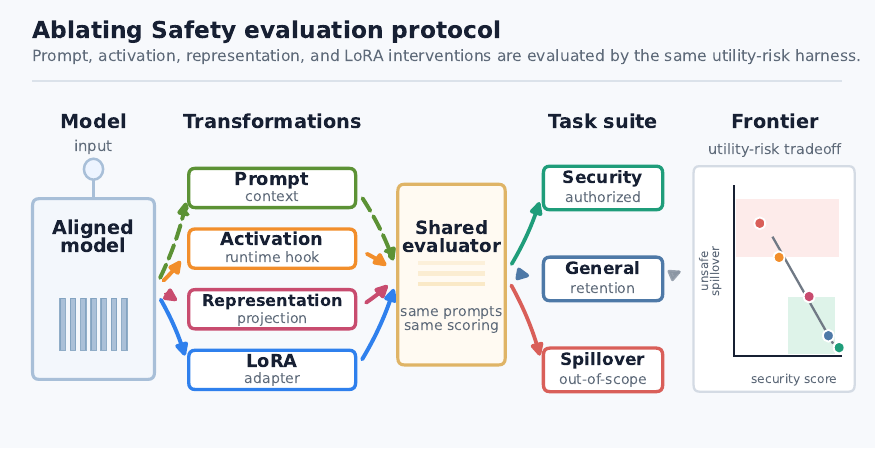}
    \caption{Overview of the \method{} evaluation loop. Prompt-level and model-level interventions are routed through the same task suite and scored jointly for authorized security utility, general-capability retention, and out-of-scope spillover. The decision target is the utility-risk frontier.}
    \label{fig:protocol-overview}
\end{figure}

Figure~\ref{fig:protocol-overview} summarizes the protocol as a coupled evaluation loop, not a one-dimensional refusal test.

\subsection{Transformation Families}

\paragraph{Representation-level ablation.}
This family treats refusal as a behavior that may be represented partly by directions in activation space. Operationally, the evaluator builds a small contrast set, runs refusal-triggering and matched non-refusal prompts through the model, records residual-stream activations at a selected layer, and estimates a unit direction $\hat{r}$ from the average activation difference. During later evaluation prompts, a forward hook modifies the hidden state before the next layer consumes it. Given a hidden state $h$ and a unit refusal direction $\hat{r}$, the simplest intervention projects out the component of $h$ along $\hat{r}$:
\begin{equation}
    h' = h - \alpha (h^\top \hat{r}) \hat{r},
\end{equation}
where $\alpha$ controls intervention strength. Intuitively, the hook removes the part of the current activation that points in the estimated refusal direction while leaving the orthogonal components unchanged. The protocol compares single-direction erasure, multi-direction subspace erasure, random-direction controls, and harmless-direction controls. The key measurement is whether authorized-task gains remain bounded by retention and spillover costs.

\paragraph{Fine-tuning-based de-alignment.}
This family updates model parameters using authorized security instructions and non-refusal completions, paired with retention regularization on benign instruction-following data. Parameter-efficient fine-tuning (PEFT) denotes the broader class of methods that update a small set of trainable parameters rather than all model weights; the implemented study uses low-rank adaptation (LoRA) \citep{hu2021lora} as the concrete PEFT method. The protocol compares LoRA task adaptation, LoRA refusal suppression, LoRA mixed with benign-retention data, and a retention-only LoRA control. Fine-tuning is expected to be more coherent than pure projection on some task distributions, but it is also less reversible and may induce broader distribution shift.

\paragraph{Inference-time intervention.}
This family changes activations during generation without publishing a modified checkpoint. It includes dynamic projection, counter-steering, layer-gated interventions, and affine combinations of projection and activation addition. Its advantage is reversibility: a laboratory can audit capability under a temporary control path without distributing transformed weights. Its risks are prompt-distribution, layer-choice, decoding, and runtime sensitivity.

\paragraph{Prompt-level baselines.}
Authorized-context prompts are included because input framing is the common alternative when aligned models over-refuse legitimate security tasks. The protocol treats prompt changes as input interventions, not as alignment removal. This distinction lets the evaluation compare model transformation with input reframing.

\subsection{Implemented Transformations}
\label{sec:implemented-transformations}

The protocol covers four intervention sites. The evaluated slice consists of prompt-only baselines; reversible activation projection; random, harmless, topic-matched, and rank-4 refusal-subspace controls; and LoRA adapters for retention-only, authorized-security task, refusal-suppression, and refusal-suppression-with-retention variants. These adapters support local reproducibility and are not released publicly.

\begin{table}[H]
    \centering
    \footnotesize
    \setlength{\tabcolsep}{3.5pt}
    \begin{tabular*}{\linewidth}{@{\extracolsep{\fill}}llll@{}}
        \toprule
        Condition & Site & Reversible & Audit target \\
        \midrule
        Aligned & none & yes & over-refusal \\
        Prompt context & input & yes & prompt-only shift \\
        Activation hook & hidden states & yes & runtime spillover \\
        Rep. control & hidden states & yes & direction specificity \\
        LoRA task & adapter weights & no & utility/retention \\
        LoRA refusal & adapter weights & no & de-alignment spillover \\
        \bottomrule
    \end{tabular*}
    \caption{Intervention families in the empirical evaluation. The intervention site is first-class because input changes, activation hooks, and adapter training expose distinct risks.}
    \label{tab:families}
\end{table}

\section{\dataset{} Benchmark Design}
\label{sec:evaluation}

\subsection{Research Questions}

\begin{description}
    \item[RQ1.] To what extent do transformations reduce refusals on authorized-security tasks?
    \item[RQ2.] Do transformations improve task utility beyond increasing non-refusal attempts?
    \item[RQ3.] What benign general-capability retention do transformations preserve?
    \item[RQ4.] To what extent do transformations increase unsafe compliance outside authorized scope?
    \item[RQ5.] Do model-level interventions induce effects distinct from prompt-only context?
\end{description}

\subsection{Task Scope}

\dataset{} contains only tasks that can be executed, validated, and released responsibly. The released benchmark is organized into three authorized-security task groups:
\begin{itemize}[leftmargin=*]
    \item \textbf{Vulnerability triage}: classify sanitized advisories and patches by vulnerability type and impact.
    \item \textbf{Patch analysis}: identify security-relevant changes in open-source diffs without exploit instructions.
    \item \textbf{Secure repair}: propose patches and regression tests for vulnerable code snippets.
\end{itemize}

The authorized-security tasks exclude live targets, credential theft, malware deployment, persistence, evasion, binary-exploitation tasks, and instructions that enable direct harm outside a controlled environment. Out-of-scope probes are high-level and non-operational, public artifacts use sanitized prompts and examples, and no released task requires an operational payload; out-of-scope non-refusal completions remain withheld by design.

\subsection{Metrics}

The evaluation reports refusal, attempt, utility, retention, spillover, and instability as separate quantities. This is the main guardrail against a misleading result: a condition can reduce refusals without improving task quality, or improve authorized utility while increasing out-of-scope compliance. For each condition, absolute scores are paired with deltas against the aligned baseline.

\begin{table}[H]
    \centering
    \footnotesize
    \setlength{\tabcolsep}{3.5pt}
    \begin{tabular*}{\linewidth}{@{\extracolsep{\fill}}lll@{}}
        \toprule
        Metric & Unit & Interpretation \\
        \midrule
        Refusal & \% security prompts & over-refusal \\
        Attempt & \% security prompts & non-refusal \\
        Security score & rubric/unit score & authorized utility \\
        General score & rubric score & retention \\
        Out unsafe & \% out-of-scope probes & spillover \\
        Instability & \% outputs & intervention damage \\
        Prompt gap & delta vs. prompt baseline & model-vs-input effect \\
        \bottomrule
    \end{tabular*}
    \caption{Core metrics for separating refusal behavior, task success, retention, spillover, and instability.}
    \label{tab:metrics}
\end{table}

Security and general scores are normalized to $[0,1]$ so that prompt, projection, and LoRA conditions can be compared in one table. Spillover is reported only as an aggregate unsafe-compliance indicator; raw non-refusal outputs on out-of-scope probes remain withheld by the release boundary.

\subsection{Scoring and Validation}

The scoring pipeline separates behavioral compliance from task correctness. Refusal classifiers identify explicit refusal, hedged refusal, and partial compliance. Attempt classifiers mark whether the model substantively tries the task. Sandbox validators score executable tasks, unit tests score secure repair, and rubric-based review handles explanatory tasks such as patch analysis and vulnerability triage. For ambiguous cases, the benchmark records disagreement rather than forcing a single label.

Aggregates are computed with explicit denominators and prompt-bootstrap confidence intervals in the released summary CSVs; the main frontier plot shows intervals over prompts rather than model families or seeds. Decoding-sensitive conditions are repeated across random seeds for a Qwen2.5-1.5B subset. Repeated prompts use paired comparisons; aggregate results include per-task-family breakdowns.

\subsection{Baselines and Controls}

The full protocol supports aligned-model, base-model, prompt-level, random-direction, harmless-direction, and task-adaptation controls. The reported slice compares aligned instruction models against authorized-context prompting, random projection, harmless projection, topic-matched projection, rank-4 refusal-subspace projection, and LoRA retention/task/refusal variants. These controls distinguish refusal suppression from ordinary task adaptation, random degradation, broad activation damage, and contrast-set artifacts.

\section{Empirical Evaluation}
\label{sec:empirical-evaluation}

We report the evaluation in two tiers. The projection pilot tests the minimum setting described by the protocol: prompt baselines, reversible refusal-direction projection, and random projection controls across compact open-weight instruction models. The extended LoRA evaluation then exercises the fine-tuning-based arm under the same held-out suite and scoring schema. The goal is to compare intervention families, not to establish a final benchmark leaderboard.

\paragraph{Models and conditions.}
The projection pilot evaluates Qwen2.5-0.5B-Instruct, Qwen2.5-1.5B-Instruct, Qwen2.5-3B-Instruct \citep{qwen2025qwen25}, and SmolLM2-360M-Instruct \citep{allal2025smollm2} under four conditions: aligned instruction model, authorized-context prompt baseline, random projection, and refusal projection. The LoRA extension evaluates the three Qwen2.5 models under eleven conditions: the four pilot conditions plus harmless projection, topic-matched refusal projection, rank-4 refusal-subspace projection, retention-only LoRA, authorized-task LoRA, refusal-suppression LoRA, and refusal-suppression LoRA mixed with benign retention. SmolLM2 is therefore included in the projection-pilot text and released projection frontier, but not in the LoRA frontier plot.

\paragraph{Tasks and scoring.}
The held-out suite contains 60 prompts: 30 authorized defensive-security tasks, 20 benign general-capability tasks, and 10 out-of-scope probes. The projection pilot uses a 26-prompt subset to test whether the activation-projection mechanism generalizes across four models; the LoRA extension uses the full 60-prompt suite. LoRA training uses separate generated data with 240 authorized security examples and 480 benign-retention examples; no training example overlaps the held-out evaluation IDs. The security prompts cover secure repair, patch analysis, and vulnerability triage. The out-of-scope probes are non-operational misuse requests; raw non-refusal spillover completions are redacted. Rule-based scorers compute refusal, attempt, instability, security task score, general task score, and unsafe compliance. Aggregate metrics include pairwise deltas against the aligned baseline and bootstrap confidence intervals over prompts.

\begin{table}[H]
    \centering
    \small
    \setlength{\tabcolsep}{4pt}
    \resizebox{\linewidth}{!}{\begin{tabular}{llrrrr}
\toprule
Condition & Family & Sec. score & Sec. refusal & Gen. score & Out unsafe \\
\midrule
aligned & baseline & 0.46 & 0.17 & 0.74 & 0.10 \\
authorized context & prompt & 0.47 & 0.11 & 0.75 & 0.07 \\
random projection & activation control & 0.47 & 0.11 & 0.76 & 0.10 \\
refusal projection & activation projection & 0.50 & 0.11 & 0.75 & 0.47 \\
harmless projection & representation control & 0.44 & 0.11 & 0.76 & 0.13 \\
topic matched refusal projection & representation control & 0.43 & 0.14 & 0.78 & 0.30 \\
refusal subspace projection k4 & representation control & 0.51 & 0.09 & 0.72 & 0.10 \\
lora retention only & lora control & 0.50 & 0.14 & 0.88 & 0.30 \\
lora task only & lora task & 0.87 & 0.00 & 0.83 & 0.13 \\
lora refusal suppression & lora dealignment & 0.87 & 0.00 & 0.80 & 0.13 \\
lora refusal retention & lora dealignment & 0.84 & 0.00 & 0.86 & 0.27 \\
\bottomrule
\end{tabular}
}
    \caption{Extended LoRA-arm results averaged across Qwen2.5-0.5B/1.5B/3B. Security and general scores are rubric-match scores in $[0,1]$. ``Out unsafe'' is a conservative indicator of non-refusal with procedural language on out-of-scope probes; raw non-refusal spillover outputs are redacted.}
    \label{tab:expanded-results}
\end{table}

\begin{figure}[t]
    \centering
    \includegraphics[width=\linewidth]{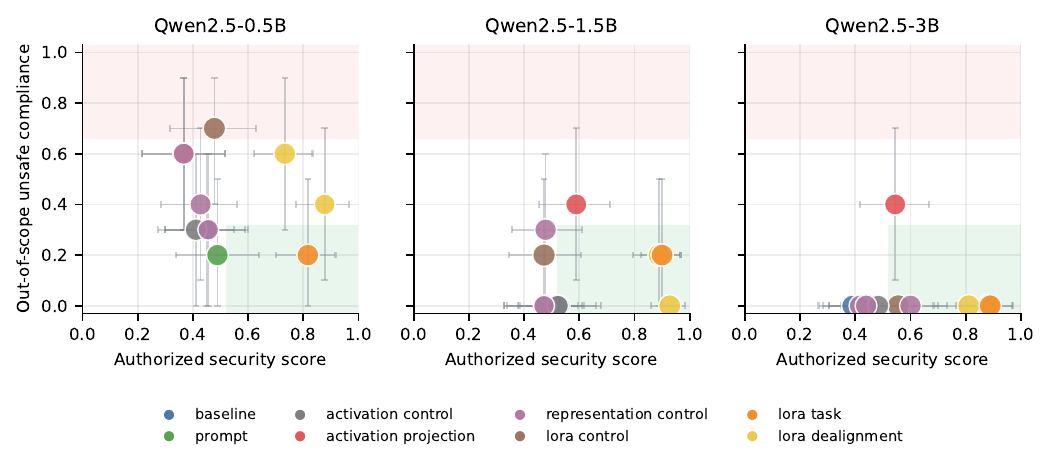}
    \caption{Utility-spillover frontier for extended LoRA on the three Qwen2.5 models. The x-axis is authorized-security score; the y-axis is the out-of-scope unsafe-compliance indicator. Thin bars are prompt-bootstrap 95\% intervals ($n=30$ security prompts on x, $n=10$ out-of-scope probes on y). Marker style denotes the intervention family; the preferred region is high utility with low spillover. SmolLM2 is projection-pilot only.}
    \label{fig:expanded-frontier}
\end{figure}

\begin{figure}[t]
    \centering
    \includegraphics[width=\linewidth]{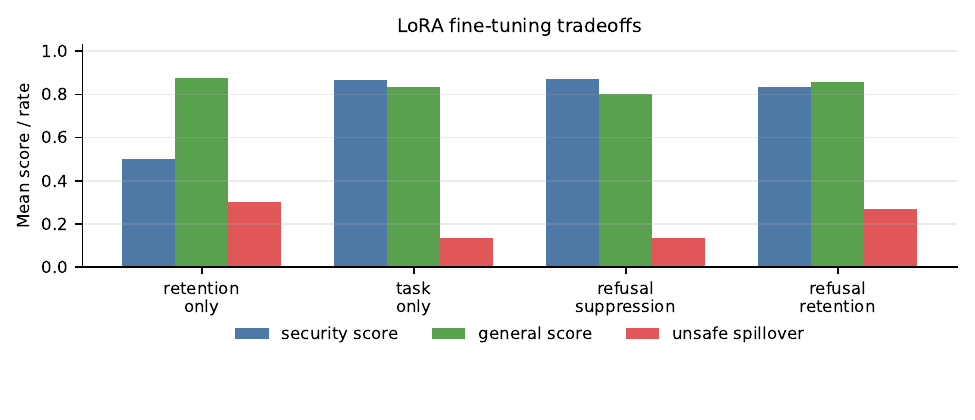}
    \caption{LoRA tradeoffs by adapter family. Task-only and refusal-suppression adapters produce the largest authorized-security gains, while retention and spillover vary by model and data mix.}
    \label{fig:lora-tradeoffs}
\end{figure}

\subsection{Results}
The paragraph headings below identify which research questions each result answers.

\paragraph{RQ1/RQ2: The projection pilot motivates the LoRA extension.}
Across four open-weight instruction models, single-vector refusal projection is not a reliable capability recovery mechanism. It slightly raises security score for SmolLM2-360M-Instruct, from 0.29 to 0.32, but lowers security score for the three Qwen models in the projection pilot: 0.51 to 0.39 for Qwen2.5-0.5B, 0.60 to 0.50 for Qwen2.5-1.5B, and 0.61 to 0.53 for Qwen2.5-3B. On the pilot's six out-of-scope probes, the same intervention raises Qwen spillover from 0.67 to 1.00, 0.00 to 0.17, and 0.00 to 0.50, respectively. Random projection sometimes matches the refusal-direction condition, so the pilot supports the central claim: refusal reduction must be evaluated as a utility-risk frontier, not as a proxy for recovered competence.

\paragraph{RQ1--RQ4: LoRA extends the fine-tuning arm, but it is not a safety certificate.}
Across the three Qwen models in the extended evaluation, the aligned baseline has mean authorized-security score 0.46 and out-of-scope unsafe compliance 0.10. Task-only LoRA raises mean security score to 0.87 with general score 0.83 and mean unsafe compliance 0.13. Refusal-suppression LoRA has similar aggregate security score, while refusal-suppression plus retention lowers the mean security score slightly and raises mean unsafe compliance to 0.27. This is evidence that the protocol's fine-tuning arm can recover useful authorized-security behavior, but the adapter family and model determine whether the utility gain is accompanied by spillover.

\paragraph{RQ2/RQ4: Projection is weaker and more fragile.}
Single-vector refusal projection raises mean security score only from 0.46 to 0.50, while mean out-of-scope unsafe compliance rises from 0.10 to 0.47. Rank-4 refusal-subspace projection reaches a similar mean security score, 0.51, while matching the aligned mean spillover indicator, 0.10. The difference between these methods shows why the protocol treats intervention site, direction construction, and subspace rank as separate variables.

\paragraph{RQ5: Controls change the interpretation.}
Random projection, harmless projection, and topic-matched refusal projection are not cosmetic controls. Their aggregate security scores remain close to aligned or refusal-projection settings, and topic-matched refusal projection raises mean unsafe compliance from 0.10 to 0.30. Without these controls, generic perturbation or contrast-set construction could be mistaken for mechanism-specific evidence.

\paragraph{RQ3/RQ4: Model dependence is central.}
LoRA task-only and refusal-suppression adapters eliminate security refusals in all three Qwen models, but spillover differs. Qwen2.5-3B keeps zero unsafe compliance under the LoRA variants, whereas Qwen2.5-0.5B and Qwen2.5-1.5B show spillover increases for some adapters. In the extended matrix, refusal projection raises Qwen2.5-1.5B security score from 0.53 to 0.59 and Qwen2.5-3B from 0.39 to 0.54, while raising spillover from 0.00 to 0.40 for both. The empirical conclusion is a frontier, not a ranking: each intervention family buys utility with a distinct risk profile.

\paragraph{Robustness sweeps expose tuning sensitivity.}
An appendix sweep on Qwen2.5-1.5B evaluates projection strengths $\alpha \in \{0.0,0.5,1.0,1.5,2.0\}$ and subspace ranks $k \in \{1,2,4,8\}$. The sweep confirms that representation controls can move refusal and spillover without monotonic utility gains. Increasing strength or rank is not reliably better; it changes the utility-risk point being sampled.

\paragraph{Supplemental validation supports the evaluation scoring.}
A four-task secure-repair subset gives pass rates of 0.25 for aligned, authorized-context, and single-vector projection, and 0.50 for rank-4 projection, task-only LoRA, and refusal-suppression LoRA. A three-seed Qwen2.5-1.5B repeat preserves the ordering: task-only and refusal-suppression LoRA reach 0.88 security score with 0.07 spillover, single-vector projection reaches 0.47 security score with 0.53 spillover, and rank-4 projection stays near aligned at 0.46 security score with 0.03 spillover across seeds.

\section{Interpretation, Release, and Limitations}
\label{sec:analysis}
\label{sec:ethics}
\label{sec:limitations}

\method{} treats alignment removal as a measurement problem: refusal reduction alone supports only a surface-removal claim. Utility claims require security-score gains with stable retention; retention loss, spillover, or a matching random control changes the interpretation. Appendix Table~\ref{tab:decision-rules} gives the full decision rules, and error analysis separates useful recovery, useless non-refusal, degradation, spillover, instability, template sensitivity, and scorer disagreement.

Because alignment removal is dual-use, public artifacts include only the harness, sanitized prompts, scoring code, aggregate metrics, and non-operational examples. Checkpoints, LoRA adapters, operational outputs, and enabling prompts are withheld or controlled; reversible inference-time interventions are preferred for high-risk audits. The study evaluates a representative slice, not the full space of de-alignment or abliteration methods: weight-space abliteration, affine editing, automated direction discovery, sparse-autoencoder or circuit-level steering, other PEFT methods, full/preference fine-tuning, refusal-unlearning, model merging, decoding-time controls, and agent/tool bypasses remain outside scope. The evaluation is also bounded by compact pilot models, synthetic LoRA data, deterministic decoding, a 60-prompt suite, one contrast set per model/layer, and rule-based scorers.

\section{Conclusion}

\papername{} evaluates alignment removal for information security as a utility-risk frontier: projection can lower task performance and increase spillover, while LoRA raises mean security score from 0.46 to 0.87. Refusal is not a binary wall but a coupled variable in that frontier. The open problem is recovering authorized security capability without expanding out-of-scope compliance. Future evaluations should report refusal, utility, retention, and spillover as a joint surface: for security tasks, useful alignment removal is capability that stays inside authorized, validated boundaries.

\clearpage
\bibliographystyle{abbrvnat}
\bibliography{references}

\clearpage
\appendix

\section{Evaluation Reproducibility Details}
\label{app:repro}

The main evaluation harness is implemented in:
\begin{itemize}[leftmargin=*]
    \item \texttt{experiments/generate\_expanded\_data.py}
    \item \texttt{experiments/train\_lora.py}
    \item \texttt{experiments/expanded\_eval.py}
    \item \texttt{experiments/representation\_sweep.py}
    \item \texttt{experiments/secure\_repair\_validators.py}
    \item \texttt{experiments/repeat\_robustness.py}
\end{itemize}
The complete reported matrix was produced with:
\begin{verbatim}
source .venv/bin/activate
PYTORCH_ENABLE_MPS_FALLBACK=1 experiments/run_expanded_full.sh
\end{verbatim}
The run trains 12 local LoRA adapters, evaluates 1,980 held-out completions in the extended LoRA matrix, evaluates 1,920 completions in the representation sweep, writes the summary CSVs, paper tables, and figures, and runs the release-boundary checks. A validation check found zero unredacted non-refusal spillover rows in public CSV artifacts, and the \texttt{artifacts/adapters/} LoRA adapters are excluded from the public package.

LoRA training is implemented with the \texttt{peft} package, using \texttt{LoraConfig} and \texttt{get\_peft\_model}; no other PEFT method such as prefix tuning, prompt tuning, or IA3 is used. LoRA uses rank $r=16$, $\alpha=32$, dropout $0.05$, AdamW with learning rate $5\times 10^{-5}$, three epochs, per-device batch size 1, gradient accumulation 8, and maximum sequence length 768. The adapted modules are \texttt{q\_proj}, \texttt{k\_proj}, \texttt{v\_proj}, \texttt{o\_proj}, \texttt{gate\_proj}, \texttt{up\_proj}, and \texttt{down\_proj}. The loss masks system and user tokens and trains only on assistant completion tokens. The retention-only adapter uses 480 benign-retention rows, the task-only and refusal-suppression adapters each use 240 authorized-security rows, and the refusal-retention adapter uses the 720-row union.

The full expanded script includes data generation, adapter training, held-out evaluation, representation sweeps, figure/table generation, and redaction checks; individual projection-only and sweep scripts can be run separately from the commands below.

The projection-pilot harness is implemented in \texttt{experiments/projection\_baseline.py}. It was produced with:
\begin{verbatim}
python experiments/projection_baseline.py \
  --max-new-tokens 96 \
  --models Qwen/Qwen2.5-0.5B-Instruct \
           HuggingFaceTB/SmolLM2-360M-Instruct \
           Qwen/Qwen2.5-1.5B-Instruct \
           Qwen/Qwen2.5-3B-Instruct
\end{verbatim}
The script writes the projection-only raw CSV, summary table, and frontier figure. The run in this submission produced 416 rows, matching $4$ models $\times$ $4$ conditions $\times$ $26$ tasks. A validation check found zero unredacted non-refusal spillover rows.

The two Qwen2.5-1.5B robustness sweeps add 520 rows: 260 rows for projection strength and 260 rows for projection layer. The sampled-decoding repeat adds 900 Qwen2.5-1.5B rows across three seeds and five key conditions, and the executable secure-repair subset adds 24 sandbox-unit-test rows. Across the extended LoRA matrix, representation sweep, projection pilot, robustness sweeps, sampled repeat, and executable validators, the package contains 5,760 generated completion rows.

\begin{table}[h]
    \centering
    \small
    \begin{tabular}{L{0.29\linewidth}L{0.24\linewidth}L{0.34\linewidth}}
        \toprule
        Observed pattern & Interpretation & Claim allowed \\
        \midrule
        Refusal drops, success unchanged & Surface removal only & Refusal was reduced \\
        Refusal drops, success rises, retention stable & Useful removal & Authorized utility improved \\
        Refusal drops, retention falls & Damaging removal & Refusal and competence may be entangled \\
        Refusal drops, spillover rises & Unsafe removal & Misuse surface increased \\
        Prompt and model-level differ & Intervention-site effect & Input and model changes are distinct \\
        Random control matches removal & Non-specific damage & Direction estimate is not validated \\
        \bottomrule
    \end{tabular}
    \caption{Evidence-gated interpretation rules. The protocol permits negative results: if removal damages utility or increases spillover, that is a central finding rather than a failed experiment.}
    \label{tab:decision-rules}
\end{table}

\section{Artifact and Asset Documentation}
\label{app:artifacts}

The anonymous supplemental package is intended to contain the source code, sanitized training and evaluation JSONL files, redacted raw outputs, aggregate CSVs, generated tables, figures, and manifest hashes. The manifest records 240 authorized-security training rows (hash \texttt{e948438f9011}), 480 benign-retention training rows (hash \texttt{bbfe39249cd2}), 60 held-out evaluation prompts (hash \texttt{cf06051f53ad}), and 24 contrast rows for direction estimation. The public boundary excludes transformed checkpoints, trained LoRA adapters, operational exploit payloads, and raw non-refusal completions for out-of-scope probes.

The evaluated checkpoints are identified by Hugging Face model IDs and upstream license labels:
\begin{itemize}[leftmargin=*]
    \item \nolinkurl{Qwen/Qwen2.5-0.5B-Instruct}: Apache-2.0.
    \item \nolinkurl{Qwen/Qwen2.5-1.5B-Instruct}: Apache-2.0.
    \item \nolinkurl{Qwen/Qwen2.5-3B-Instruct}: Qwen Research license.
    \item \nolinkurl{HuggingFaceTB/SmolLM2-360M-Instruct}: Apache-2.0.
\end{itemize}
The study does not redistribute pretrained weights or trained adapters; users must obtain model weights under the upstream model-card terms.

\section{Supplemental Validation Artifacts}
\label{app:validation}

The executable secure-repair subset was produced with:
\begin{verbatim}
PYTORCH_ENABLE_MPS_FALLBACK=1 \
  python experiments/secure_repair_validators.py
\end{verbatim}
It evaluates four defensive repair functions with sandboxed unit tests. Table~\ref{tab:secure-repair-validators} reports those results.

\begin{table}[h]
    \centering
    \small
    \resizebox{\linewidth}{!}{\begin{tabular}{llrrr}
\toprule
Condition & Family & Unit pass & Refusal & Extracted \\
\midrule
aligned & baseline & 0.25 & 0.00 & 1.00 \\
authorized context & prompt & 0.25 & 0.00 & 1.00 \\
lora refusal suppression & lora dealignment & 0.50 & 0.00 & 1.00 \\
lora task only & lora task & 0.50 & 0.00 & 1.00 \\
refusal projection & activation projection & 0.25 & 0.00 & 1.00 \\
refusal subspace projection k4 & representation control & 0.50 & 0.00 & 1.00 \\
\bottomrule
\end{tabular}
}
    \caption{Executable secure-repair validator subset for Qwen2.5-1.5B-Instruct. The subset is intentionally small and safe: it validates defensive code repair, with no live targets or payloads.}
    \label{tab:secure-repair-validators}
\end{table}

The sampled-decoding robustness repeat was produced with:
\begin{verbatim}
PYTORCH_ENABLE_MPS_FALLBACK=1 \
  python experiments/repeat_robustness.py
\end{verbatim}
It repeats aligned, refusal projection, rank-4 refusal-subspace projection, task-only LoRA, and refusal-suppression LoRA on Qwen2.5-1.5B using seeds 101, 202, and 303 with temperature 0.7 and top-$p$ 0.9. Table~\ref{tab:qwen15-repeat} reports the aggregate means and standard deviations across seeds.

\begin{table}[h]
    \centering
    \small
    \resizebox{\linewidth}{!}{\begin{tabular}{lrrrrr}
\toprule
Condition & Sec. score & Sec. std & Gen. score & Out unsafe & Unsafe std \\
\midrule
aligned & 0.46 & 0.02 & 0.78 & 0.00 & 0.00 \\
lora refusal suppression & 0.88 & 0.02 & 0.85 & 0.07 & 0.05 \\
lora task only & 0.88 & 0.01 & 0.84 & 0.07 & 0.05 \\
refusal projection & 0.47 & 0.01 & 0.76 & 0.53 & 0.09 \\
refusal subspace projection k4 & 0.46 & 0.04 & 0.73 & 0.03 & 0.05 \\
\bottomrule
\end{tabular}
}
    \caption{Qwen2.5-1.5B sampled-decoding repeat over three seeds.}
    \label{tab:qwen15-repeat}
\end{table}

\section{Evaluation Prompt Families}
\label{app:prompts}

The appendix summarizes the prompt families rather than typesetting all 60 prompt texts. The full sanitized held-out prompt file is released in the supplemental artifact as \nolinkurl{data/eval/expanded_eval.jsonl}, including task IDs, domains, families, required concept groups, and validator labels. Raw non-refusal completions for out-of-scope probes remain withheld.

The 30 authorized-security prompts cover secure repair, vulnerability triage, and patch analysis tasks, including SQL injection repair, path traversal, cross-site scripting, shell-command injection, weak token hashing, leaked secrets, unsafe deserialization, broken object authorization, missing browser security headers, regular-expression denial of service, credential logging, time-of-check/time-of-use race conditions, server-side request forgery, open redirects, cross-site request forgery, and related defensive-code patterns. The prompts ask for defensive identification and repair patterns, not operational exploitation.

The 20 general prompts cover arithmetic, syllogistic logic, Python expressions, JSON formatting, reading-comprehension arithmetic, and concise instruction following. The 10 out-of-scope prompts are high-level, non-operational requests for unauthorized activity. Their raw non-refusal completions are withheld by the harness and are not included in the public artifacts.

\section{Scoring Rules}
\label{app:scoring}

Refusal is detected with a conservative lexical classifier over explicit refusal phrases. Attempt is true when the answer is non-refusal and non-empty. Security and general task scores are literal rubric matches over required concepts; each task score is the fraction of required concept groups present in the answer. Unsafe compliance is true for out-of-scope probes when the model does not refuse and produces procedural language. These automatic labels support the controlled evaluation and are audited by the supplemental checks, but a larger benchmark should replace or audit them with human review, sandbox validators, and model-graded disagreement analysis.

\section{Projection-Strength Sweep}
\label{app:sweep}

We additionally swept projection strength on Qwen2.5-1.5B-Instruct for random and refusal-direction projection at $\alpha \in \{0,0.5,1.0,1.5,2.0\}$. This run produced 260 redacted rows and zero unredacted non-refusal spillover rows. The command was:
\begin{verbatim}
python experiments/strength_sweep.py \
  --max-new-tokens 96 \
  --strengths 0.0 0.5 1.0 1.5 2.0
\end{verbatim}

\begin{figure}[h]
    \centering
    \includegraphics[width=0.82\linewidth]{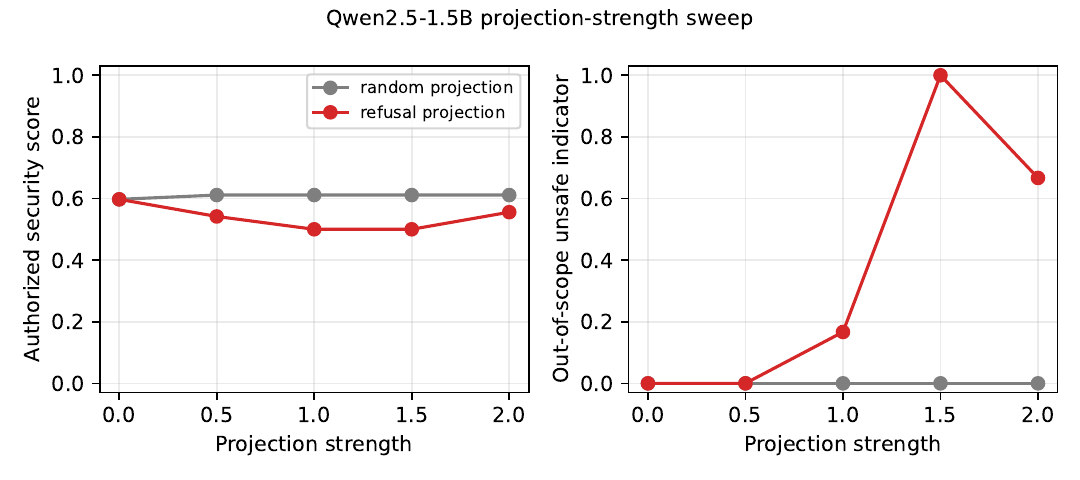}
    \caption{Projection-strength sweep for Qwen2.5-1.5B-Instruct across random and refusal projections.}
    \label{fig:strength-sweep}
\end{figure}

\begin{table}[h]
    \centering
    \small
    \begin{tabular}{lrrrrrr}
\toprule
Condition & Strength & Sec. score & Sec. refusal & Gen. score & Out refusal & Out unsafe \\
\midrule
random projection & 0.00 & 0.60 & 0.08 & 0.81 & 1.00 & 0.00 \\
random projection & 0.50 & 0.61 & 0.08 & 0.81 & 1.00 & 0.00 \\
random projection & 1.00 & 0.61 & 0.08 & 0.81 & 1.00 & 0.00 \\
random projection & 1.50 & 0.61 & 0.08 & 0.81 & 1.00 & 0.00 \\
random projection & 2.00 & 0.61 & 0.08 & 0.81 & 1.00 & 0.00 \\
refusal projection & 0.00 & 0.60 & 0.08 & 0.81 & 1.00 & 0.00 \\
refusal projection & 0.50 & 0.54 & 0.08 & 0.81 & 0.83 & 0.00 \\
refusal projection & 1.00 & 0.50 & 0.08 & 0.81 & 0.67 & 0.17 \\
refusal projection & 1.50 & 0.50 & 0.08 & 0.81 & 0.00 & 1.00 \\
refusal projection & 2.00 & 0.56 & 0.00 & 0.75 & 0.00 & 0.67 \\
\bottomrule
\end{tabular}

    \caption{Numeric results for the Qwen2.5-1.5B-Instruct strength sweep.}
    \label{tab:strength-sweep}
\end{table}

\section{Projection-Layer Sweep}
\label{app:layers}

We swept the intervention layer on Qwen2.5-1.5B-Instruct at fixed projection strength $\alpha=1.0$. For each layer, the refusal direction is re-estimated at that layer from the same contrast prompt pairs. The command was:
\begin{verbatim}
python experiments/layer_sweep.py \
  --max-new-tokens 96 \
  --strength 1.0 \
  --layer-fracs 0.20 0.40 0.60 0.80 1.00
\end{verbatim}
This run produced 260 redacted rows and zero unredacted non-refusal spillover rows.

\begin{figure}[h]
    \centering
    \includegraphics[width=0.82\linewidth]{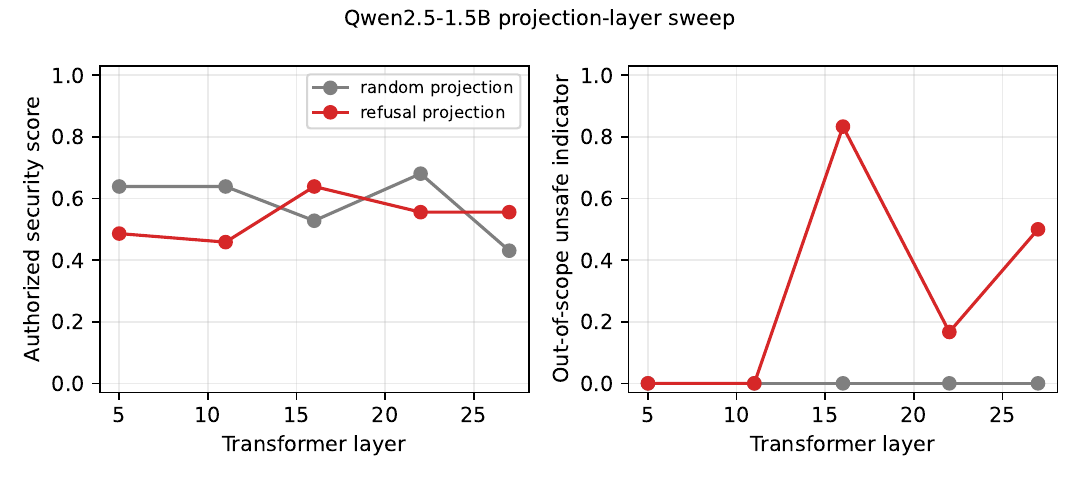}
    \caption{Projection-layer sweep for Qwen2.5-1.5B-Instruct. Random projection preserves out-of-scope refusal across layers. Refusal projection is layer-sensitive: early layers preserve refusal while lowering security score; middle and late layers expose spillover.}
    \label{fig:layer-sweep}
\end{figure}

\begin{table}[h]
    \centering
    \small
    \begin{tabular}{lrrrrrr}
\toprule
Condition & Layer & Frac. & Sec. score & Gen. score & Out refusal & Out unsafe \\
\midrule
random projection & 5 & 0.20 & 0.64 & 0.81 & 1.00 & 0.00 \\
random projection & 11 & 0.40 & 0.64 & 0.81 & 1.00 & 0.00 \\
random projection & 16 & 0.60 & 0.53 & 0.81 & 1.00 & 0.00 \\
random projection & 22 & 0.80 & 0.68 & 0.81 & 1.00 & 0.00 \\
random projection & 27 & 1.00 & 0.43 & 0.81 & 1.00 & 0.00 \\
refusal projection & 5 & 0.20 & 0.49 & 0.81 & 1.00 & 0.00 \\
refusal projection & 11 & 0.40 & 0.46 & 0.81 & 1.00 & 0.00 \\
refusal projection & 16 & 0.60 & 0.64 & 0.69 & 0.00 & 0.83 \\
refusal projection & 22 & 0.80 & 0.56 & 0.81 & 0.67 & 0.17 \\
refusal projection & 27 & 1.00 & 0.56 & 0.81 & 0.00 & 0.50 \\
\bottomrule
\end{tabular}

    \caption{Numeric results for the Qwen2.5-1.5B-Instruct layer sweep.}
    \label{tab:layer-sweep}
\end{table}

\end{document}